\documentclass[12pt]{article}
\usepackage{amsmath,amssymb}
\usepackage{graphicx,color}
\numberwithin{equation}{section}
\usepackage{cite}
\usepackage{bm}
\usepackage{dcolumn}
\newcommand{\be}{\begin{equation}}
\newcommand{\bea}{\begin{eqnarray}}
\newcommand{\eea}{\end{eqnarray}}
\newcommand{\ba}{\begin{array}}
\newcommand{\ea}{\end{array}}
\newcommand{\ee}{\end{equation}}

\expandafter\ifx\csname mathbbm\endcsname\relax

\else

\fi \textheight 22cm \textwidth 15cm \topmargin 1mm
\def\p{\partial}
\begin{document}

\begin{titlepage}
\hfill
\vbox{
    \halign{#\hfil         \cr
           IPM/P-2008/065 \cr
                      } 
      }  
\vspace*{20mm}
\begin{center}
{\Large {\bf 2D Gravity on $AdS_2$ with Chern-Simons Corrections}\\
}

\vspace*{15mm}
\vspace*{1mm}
{Mohsen Alishahiha$^a$, Reza Fareghbal$^a$  and  Amir E. Mosaffa$^{a,b}$}

 \vspace*{1cm}

{\it ${}^a$ School of physics, Institute for Research in Fundamental Sciences (IPM)\\
P.O. Box 19395-5531, Tehran, Iran \\ }

\vspace*{.4cm}

{\it ${}^b$ Department of Physics, Sharif University of Technology \\
P.O. Box 11365-9161, Tehran, Iran}

\vspace*{2cm}
\end{center}

\begin{abstract}

We study 2D Maxwell-dilaton gravity with higher order corrections
given by the Chern-Simons term. The model admits three distinctive
$AdS_2$ vacuum solutions. By making use of the entropy function
formalism we find the entropy of the solutions which is corrected
due to the presence of the Chern-Simons term. We observe that the
form of the correction depends not only on the coefficient of the
Chern-Simons term, but also on the sign of the electric charge;
pointing toward the chiral nature of the dual CFT. Using the
asymptotic symmetry of the theory as well as requiring a consistent
picture we can find the central charge and the level of $U(1)$
current. Upon uplifting the solutions to three dimensions we get
purely geometric solutions which will be either $AdS_3$ or warped
$AdS_3$ with an identification.

\end{abstract}

\vspace{1.5cm}
 \begin{center}
 \it{Dedicated to Hessamaddin  Arfaei on the occasion of his 60th birthday}
 \end{center}
\end{titlepage}

\section{Introduction}

2D quantum gravity on $AdS_2$ geometry is important due to its
essential role in the context of black hole physics. Indeed the
$AdS_2$ geometry is the factor which appears in the near horizon
geometry of extremal black holes in any dimension. Therefore
understanding quantum gravity on $AdS_2$ might ultimately help us
understand the origin of the black hole entropy in other
dimensions.

The main problem which prevents us to explain quantum gravity on
$AdS_2$ geometry is the fact that it is not quite clear what it
actually means. Indeed this is the case for any dimension. An
attempt to understand, or in better words, to make sense of
quantum gravity in three dimensions has been made by Witten in
\cite{witten} where it was argued that 3D quantum gravity makes
sense only on $AdS_3$. The main reason supporting the argument is
due to the existence of non-trivial three dimensional black holes,
BTZ solutions, which carry non-zero entropy \cite{Banados:1992wn}.
Being an AdS background it is natural to define the quantum
gravity in terms of  the dual CFT via AdS/CFT correspondence
\cite{Maldacena:1997re}.

In 2D Maxwell-dilaton gravity there are several classical
solutions with non-zero entropy which may be interpreted as 2D
extremal black holes. Therefore we would expect to have
non-trivial 2D quantum gravity on $AdS_2$ geometry. Following the
idea explored in \cite{witten} one may suspect that quantum
gravity on $AdS_2$ can be defined via its CFT dual. We note,
however, that although AdS$_{d+1}$/CFT$_{d}$ correspondence has
been understood for $d\geq 2$ mainly due to explicit examples,
little has been known for the case of $d=1$ and indeed it remains
enigmatic. Nevertheless there are several attempts to explore
$AdS_2/CFT_1$ correspondence, including
\cite{{Strominger:1998yg},{Cadoni:1999ja},{NavarroSalas:1999up},
{Park:1999hs},{Cadoni:2000ah},
{Cadoni:2000gm},{Astorino:2002bj},{Kim:1998wy},{Leiva:2003kd},{Hyun:2007ii},{Correa:2008bi},{Kang:2004js},
 {HS},{Sen:2008yk},{Alishahiha:2008tv},{Gupta:2008ki},{Cadoni:2008mw},
{Castro:2008ms},{Sen:2008vm},{Morita:2008qn}}.

The aim of the present article is to further study 2D gravity on
$AdS_2$ along the recent studies
\cite{{HS},{Alishahiha:2008tv},{Castro:2008ms}} where 2D
Maxwell-dilaton gravity has been considered. To have  consistent
boundary conditions it was shown in \cite{HS} that the asymptotic
symmetry of the model is generated by a twisted energy momentum
tensor whose central charge is non-zero. This central charge along
with eigenvalue of $L_0$ of the dual CFT can be used to {\it
consistently} reproduce the entropy of the bulk gravity via the
Cardy formula. This was taken as an evidence that the CFT dual to
gravity on $AdS_2$ should be {\it a chiral half of a 2D CFT}.

To elaborate the above statement we will consider 2D
Maxwell-dilaton gravity in the presence of higher order
corrections given by 2D Maxwell-gravitational Chern-Simons term.
To have consistent boundary conditions one needs to work with the
twisted energy momentum tensor, though in this case due to the
presence of the Chern-Simons term the corresponding central charge
gets a correction. An important observation is that the correction
not only depends on the coefficient of the Chern-Simons term, but
also it is sensible to the sign of the electric charge. The sign
dependent effect should, indeed, be associated with the fact that
the dual theory should be a chiral half of a 2D CFT.

To study the vacuum solutions of the model we should solve the
equations of motion with a constant dilaton. Equivalently, we may
utilize the entropy function formalism \cite{SEN} by which we are
also able to find the entropy of the corresponding solutions. From
the equations of motion we find three distinctive $AdS_2$ vacuum
solutions. Using the asymptotic symmetry of the theory together
with requiring to have a consistent picture we will be able to
read the central charge of the corresponding solutions as well as
the level of $U(1)$ current.

The 2D solutions may be uplifted to three dimensions. The obtained
3D solutions are purely geometric solutions that will be either
$AdS_3$ or warped $AdS_3$ with an identification. The warped $AdS_3$
solution has recently been studied in \cite{strominger2} (see also
\cite{{compere},{Anninos:2008qb},{Carlip:2008eq},{Gibbons:2008vi}}).

The paper is organized as follows. In the next section we will
introduce our model where we apply the entropy function formalism
to find the vacuum solutions as well as their entropy. Re-writing
the entropy in a suggestive form, we will give an expression for
the corrected central charge. In section 3 requiring to have
consistent boundary conditions  we will find the asymptotic
symmetry of the theory which can be used to read the central
charge. In section 4 we uplift the 2D solutions to three
dimensions which may be compared with 3D solutions in
\cite{strominger2}. The last section is devoted to discussions.

\section{2D Maxwell-dilaton gravity with Chern-Simons term}

Let us consider 2D Maxwell-dilaton gravity with the action
\be\label{action1} S=S_{EH}+S_{CS} \ee where $S_{EH}$ is the
Einstein-Hilbert action \bea\label{actions} S&=&\frac{1}{8G}\int
d^2x\sqrt{-g}\;
e^{\phi}\left(R+2\partial_\mu\phi\,\partial^\mu\phi+\frac{2}{l^2}e^{2\phi}-\frac{l^2}{4}F_{\mu\nu}F^{\mu\nu}\right),
\eea and $S_{CS}$ is the two dimensional Chern-Simons term given by
\be S_{cs}=-\frac{1}{32G\mu}\int
d^2x\left(lR\epsilon^{\mu\nu}F_{\mu\nu}+l^3
\epsilon^{\mu\nu}F_{\mu\rho}F^{\rho\delta} F_{\delta\nu}\right). \ee
The action $S_{EH}$ can actually be obtained from the 3D pure
gravity with cosmological constant by reducing to two dimensions
along an $S^1$ \cite{Strominger:1998yg}. Similarly one may start
from 3D gravitational Chern-Simons term and reduce along an $S^1$ to
arrive at the 2D Chern-Simons $S_{cs}$
\cite{{Guralnik:2003we},{Grumiller:2003ad}}. From three dimensional
point of view these actions have been used to study the entropy of
extremal black holes in the presence of higher order corrections
(see for example \cite{sen2}).

The aim of this section is to study the vacuum solutions of the
model given by the action \eqref{action1} which can be obtained by
solving its equations of motion. In fact, setting
$F_{\mu\nu}=\sqrt{-g}\epsilon_{\mu\nu} F$, the equations of motion
are given by
\begin{equation}\label{eq metric}
\begin{split}
&g_{\mu\nu}\left(\nabla^2e^\phi+\frac{1}{l^2}\,e^{3\phi}-\frac{l^2F^2}{4}\,e^\phi
+e^\phi\partial_\mu\phi\,\partial^\mu\phi\right)-\nabla_\mu\nabla_\nu
e^\phi-2e^{\phi}\partial_\mu\phi\partial_\nu\phi \cr
&-\frac{l}{2\mu}\left[g_{\mu\nu}\left(
\nabla^2F-{l^2F^3}-\frac{R}{2}F\right)-\nabla_\mu\nabla_\nu F
\right]=0
\end{split}
\end{equation}
\begin{equation}
\label{eq scalar}
    R+\frac{6}{l^2}e^{2\phi}+\frac{l^2}{2}F^2+2e^\phi\partial_\alpha\phi\partial^\alpha\phi-4\nabla^2e^\phi=0,\;\;\;
\epsilon^{\mu\nu}\partial_\mu\bigg{(}e^\phi
F+\frac{1}{2\mu l}(R+3l^2F^2)\bigg{)}=0
\end{equation}
It is useful to work with trace and traceless parts of the equation \eqref{eq metric}
\begin{equation}\label{trace part}
   \nabla^2e^\phi+\frac{2}{l^2}\,e^{3\phi}-\frac{l^2F^2}{2}\,e^\phi
=\frac{l}{2\mu}\left( \nabla^2F-2{l^2F^3}-{R}F\right)
\end{equation}
\begin{equation}\label{non trace part}
    g_{\mu\nu}(\nabla^2e^\phi+2e^\phi\partial_\alpha\phi\,\partial^\alpha\phi)-2(
    \partial_\mu\phi\,\partial_\nu\phi+\nabla_\mu\nabla_\nu
e^\phi)=\frac{l}{2\mu}\left(g_{\mu\nu}\nabla^2F-2\nabla_\mu\nabla_\nu
F\right)
\end{equation}

This model admits $AdS_2$ vacuum solutions. To find them we should
look for solutions with a constant dilaton. In this case one has
 \begin{equation}
\frac{2}{l^2}\,e^{3\phi}-\frac{l^2F^2}{2}\,e^\phi
=-\frac{l}{2\mu}\left(2{l^2F^3}+{R}F\right),\;\;\;\;
R+\frac{6}{l^2}e^{2\phi}+\frac{l^2}{2}F^2=0,
\end{equation}
which, for a given gauge field, can be solved to find the constant
dilaton. Indeed, these equations reduce to the following equation
for dilaton \be \left(e^{\phi}-\frac{3l}{2\mu}F\right)
\left(\frac{2}{l^2}e^{2\phi}-\frac{l^2}{2}F^2\right)=0. \ee
Therefore, for arbitrary $\mu$ and $l$, the model may have three
different vacuum solutions with constant dilaton given by \be
e^{\phi}=\pm\frac{l^2}{2}F,\;\;\;\;\;\;\;\;\;e^{\phi}=\frac{3}{\mu
l}\frac{l^2}{2}F. \ee It is worth noting that, as it is evident
from the above expressions, in the special case of $\mu l=3$ the
third solution degenerates with the first one (the positive sign
above). We will come back to this point later.

To find the whole solutions we need to plug these expressions for
the dilaton into the equations of motion and solved for metric and
gauge field. Equivalently, since the solutions we are looking for
are $AdS_2$, one may utilize the entropy function formalism
\cite{SEN}. An advantage of the entropy function formalism is that
with this method  we can not only find the solutions, but also we
can read the entropy of the corresponding solutions.

To proceed let us start from an ansatz preserving $SO(1,2)$
symmetry of the $AdS_2$ solution
 \be
 ds^2=v(-r^2dt^2+\frac{dr^2}{r^2}),\;\;\;\;\;\;e^\phi=u,
 \;\;\;\;\;F_{01}=\frac{e}{l^2}.
 \ee
 The entropy function is given by
 \be {\cal
 E}=2\pi[qe-f(e,v,u)]
 \ee
 where $f(e,v,u)$ is the Lagrangian
density evaluated for the above ansatz. The parameters $e,v$ and
$u$ can be obtained by extremizing the entropy function with
respect to them. Then the entropy is given by the value of the
entropy function evaluated at the extremum.

Using the above ansatz, the entropy function for the action
\eqref{action1} reads
 \be
 {\cal E}=2\pi\left\{qe-\frac{1}{8G}\left[-2u+\frac{2u^3v}{l^2}+\frac{e^2u}{2vl^2}+\frac{1}{2\mu}
 \left(\frac{2e}{vl}-\frac{e^3}{v^2 l^3}\right)\right]\right\}
 \ee
Extremizing the entropy function with respect to the parameters
$v,u$ and $e$, for generic $\mu$ and $l$  we find three different
solutions
 \bea\label{solutions} 1:&&v=\frac{1+1/\mu
 l}{-16Gq},\;\;\;e^{2\phi}=\frac{-4Gql^2}{1+1/\mu l},\;\;\;\;\;\;\;
 \frac{e}{l}=- \sqrt{\frac{1+1/\mu l}{-16qG}},\;\;\;\;\;\;\;\;\;q<
 0,\cr &&\cr 2:&&v=\frac{1-1/\mu
 l}{16Gq},\;\;\;e^{2\phi}=\frac{4Gql^2}{1-1/\mu l},\;\;\;\;\;\;
 \frac{e}{l}= \sqrt{\frac{1-1/\mu
 l}{16qG}},\;\;\;\;\;\;\;\;\;\;\;\;\;q> 0,\cr &&\cr
 3:&&v=\frac{1}{8Gq\mu l},\;\;\;\;\;\;e^{2\phi}=\frac{72Gq\mu
 l^3}{\mu^2l^2+27},\;\;\;\;\;
 \frac{e}{l}=\sqrt{\frac{\mu l}{2Gq(\mu^2l^2+27)}},\;\;\;q>0.
 \eea
The entropy of the corresponding solutions written in a suggestive form is given by
 \bea
 1:&&S=2\pi\sqrt{\frac{-ql^2}{6}\frac{3}{2G}(1+\frac{1}{\mu l})},\cr
 2:&&S=2\pi\sqrt{\frac{ql^2}{6}\frac{3}{2G}(1-\frac{1}{\mu l})},\cr
 3:&&S=2\pi \sqrt{\frac{ql^2}{6}\frac{12\mu l}{G(\mu^2l^2+27)}},
 \eea
which may be compared with the Cardy formula for the entropy
$S=2\pi\sqrt{\frac{L_0}{6}c}$. Following the general philosophy of
the AdS/CFT correspondence \cite{Maldacena:1997re} if we assume
that the 2D gravity on the $AdS_2$ solutions \eqref{solutions} has
a dual CFT, it is then natural to identify $ql^2$ with the
eigenvalue of $L_0$ of the dual CFT. Then the central charges of
the corresponding CFTs read
 \be\label{central charges1}
 1:\;\;c_R=\frac{3}{2G}(1+\frac{1}{\mu l}),\;\;\;\;\;\;2:\;\;
 c_L=\frac{3}{2G}(1-\frac{1}{\mu l}),\;\;\;\;\;\;
 3:\;\;c_L=\frac{12\mu l}{G(\mu^2 l^2+27)}.
 \ee
If correct, this means that the 2D Maxwell-dilaton gravity on
$AdS_2$ background \eqref{solutions} is dual to a chiral half of a
2D CFT characterized by the above central charges. We note,
however, that since the identification of $L_0=ql^2$ was
speculative, the above presentation cannot be considered as an
argument supporting $AdS_2/CFT_1$ correspondence. The best we can
say is that as far as the entropy is concerned, with this
identification, the picture seems self consistent. It is worth
noting that for the case of $\mu\rightarrow \infty$ where the
effect of the Chern-Simons term is zero, we recove the known
results in the literature (see for example
\cite{{Alishahiha:2008tv},{Castro:2008ms}}) legitimating our
identifications. In the next section we will present another
calculation supporting self consistency of the above picture.

The indices $L,R$ in equations \eqref{central charges1} refer to
the fact that, depending on the sign of $q$, the dual chiral CFT
is left or right handed. Moreover, as we have already mentioned
$\mu l=3$ is a special point. Indeed at this point the solution
(3) degenerates with solution (2) where we get \be
c_R=\frac{2}{G},\;\;\;\;\;\;\;c_L=\frac{1}{G}. \ee Another
interesting point  is $\mu l=\pm 1$ where we have two solutions
with following central charges \be
c_R=\frac{3}{G},\;\;\;\;\;\;\;c_L=\frac{3}{7G}, \ee
In section 4 we will compare
these results with the solutions of 3D gravity coupled to Chern-Simons term.

\section{Asymptotic symmetry and central charge}

In this section we closely follow \cite{HS} to study 2D
Maxwell-dilaton quantum gravity on the three different $AdS_2$
backgrounds in \eqref{solutions}. We will see in order to have
consistent boundary conditions the usual conformal
diffeomorphisim, generated by the energy momentum tensor of
\eqref{action1}, must be accompanied by a $U(1)$ gauge
transformation. As a result we will have to work with a twisted
energy momentum whose central charge is non-zero \cite{HS}. We
note, however, that although we would expect to get three
different central charges for three solutions in
\eqref{solutions}, since all the solutions are obtained from the
same action, \eqref{action1}, the procedure as well as the
expressions for different quantities must be universal.

To proceed we note that the $AdS_2$ vacuum solutions, setting $r=\frac{1}{\sigma}$,
can be recast to the following form
 \be
 ds^2=-4v\frac{dt^+dt^-}{(t^+-t^-)^2},\;\;\;\;\;\;\;\;
 A_{\pm}=-\frac{e}{2\sigma l^2},\;\;\;\;\;\;\;\;\;u=\eta={\rm constant},
 \ee
where $t^{\pm}=t\pm\sigma$ and $v,e,u$ are given in \eqref{solutions}.

Now the aim is to study 2D quantum gravity whose vacuum is given
by either of the above solutions. To do so, we first need to
understand the action of the conformal group on the theory. For
this purpose, following the standard procedure in 2D CFT, we
choose an appropriate gauge for the metric and the gauge field.
For the metric we choose the conformal gauge \be
 ds^2=-e^{2\rho} dt^+dt^-
\ee
and for the gauge field the Lorentz gauge
\be
\p_+A_-+\p_-A_+=0
\ee
In this gauge the gauge field can be written as
$A_{\pm}=\pm\p_{\pm}a$, for a scalar field $a$, such that
$F_{+-}=-2\p_+\p_-a$. Our gauge choice fixes the coordinates and $U(1)$ gauge field up to residual
conformal and gauge transformations generated by
 \be t^{\pm}\rightarrow t^{\pm}+\zeta^{\pm}(t^{\pm})\ , \ \ \ \ \ \
 \ \ \ a\rightarrow a+\theta(t^+)-\tilde{\theta}(t^-)
 \ee
In this gauge the action \eqref{action1} reads
 \bea
 S_{GF}&=&\frac{1}{4G}\int\ d^2t
 \ [-2\p_-\eta\p_+\rho+\frac{e^{2\rho}}{2l^2}\
 \eta^3-2\frac{\p_+\eta\p_-\eta}{\eta}+\frac{l^2}{2}\ e^{-2\rho}\eta\ (F_{+-})^2]
 \cr \nonumber\\
 &-&\frac{l}{4G\mu}\int\ d^2t\ [2e^{-2\rho}\p_+\p_-\rho\
 F_{+-}+l^2e^{-4\rho}(F_{+-})^3]
 \eea
This action should be accompanied by the equations of motion for
the fields that have been fixed by the gauge choice. These show up
as the following constraints
 \bea
 \frac{2}{\sqrt{-g}}\frac{\delta S}{\delta
 g^{\pm\pm}}\equiv
 T_{\pm\pm}&=&\frac{1}{4G}\left(-2\p_{\pm}\rho\p_{\pm}\eta+\p_{\pm}\p_{\pm}\eta+2
 \frac{\p_{\pm}\eta\p_{\pm}\eta}{\eta}\right)\cr &&\cr &+&
 \frac{l}{8G\mu}\bigg(-2\p_{\pm}\rho\p_{\pm}F+\p_{\pm}\p_{\pm}F\bigg)=0
 \eea

 \be\label{Gcons}
 -\frac{\delta S}{\delta A_{\pm}}\equiv G_{\mp}=\pm
 \frac{l^2}{8G}\p_{\mp}(\eta F)+j_{\mp}=0
 \ee
where
\be\label{div}
 j_{\pm}=\mp\frac{l}{16G\mu}\p_{\pm}(8e^{-2\rho}\p_+\p_-\rho+3l^2F^2),\;\;\;\;\;\;{\rm with}\;\;\;\p_-j_++\p_+j_-=0.
 \ee
On the other hand since we require no current flow out of the
boundary one should impose the condition
 $j_\sigma|_{\sigma=0}=0$ which, using the equation \eqref{Gcons}, we find
 \be\label{FF}
j_{\sigma}= j_+-j_-=-\frac{l^2}{8G}\  \p_t(\eta F)=0, \;\;\;\;{\rm at}\;\;\;\sigma=0.
 \ee
As a result, the boundary terms in the variation of the action
will vanish if\footnote{The constraints \eqref{Gcons} together
with \eqref{div} and the boundary condition
$j_{\sigma}|_{\sigma=0}=0$, completely determine $j$. Indeed from the variation of the action with
respect to $a$ we find a boundary term as $\partial_+(\eta F)\;\delta a$ which must be zero at the
boundary. On the other hand due to \eqref{FF} we are led to $\delta a|_{\sigma=0}=0$. This forces
a Dirichlet boundary condition for the field $a$.}
 \be\label{boundary}
 \p_t a|_{\sigma=0}=A_{\sigma}|_{\sigma=0}=0
 \ee
In general the boundary condition \eqref{boundary} is not
preserved by the remaining allowed diffeomorphisms and hence the
coordinate transformations should be accompanied by appropriate
gauge transformations \cite{HS} \be\label{gauge}
 \theta(t^+)=\frac{e}{2 l^2}\p_+\zeta^+\ ,\ \ \ \ \ \ \ \
 \tilde{\theta}(t^-)=-\frac{e}{2l^2}\p_-\zeta^-\ .
 \ee
Therefore the improved conformal transformations are generated by
the twisted energy momentum tensor
 \be
 {\tilde T}_{\pm\pm}=T_{\pm\pm}\mp \frac{e}{2l^2}\partial_\pm {\cal G}_\pm,
 \ee
where ${\cal G}_{\pm}$ is the current that generates the gauge
transformations \eqref{gauge}. Denoting by $k$ the level of $U(1)$
current which parameterizes the gauge anomaly due to the Schwinger
term, the central charge of the model reads
 \be\label{central}
 c=3k\frac{e^2}{l^4}.
 \ee
The main challenge is to find the level of $U(1)$, $k$. In general
it can be obtained by making use of the anomaly calculations
\cite{{Manton:1985jm},{Heinzl:1991vd}}. We note, however, that it
can be fixed using the known solutions. In particular for the case
of $\mu \rightarrow \infty$ where the theory is given by the first
action in \eqref{action1}, the central charge is found to be
$\frac{3}{2G}$ \cite{{Alishahiha:2008tv},{Castro:2008ms}}.
Equating this value with the central charge in \eqref{central} and
using the first or second solution in \eqref{solutions} in the
limit of $\mu \rightarrow \infty$ one finds $k=8|q| l^2$. Plugging
this back into the equation \eqref{central} we get
 \be
 1)\;\;c_R=\frac{3}{2G}(1+\frac{1}{\mu
 l}),\;\;\;\;\;\;2)\;\;c_L=\frac{3}{2G}(1-\frac{1}{\mu
 l}),\;\;\;\;\;\; 3)\;\;c_L=\frac{12\mu l}{G(\mu^2 l^2+27)},
 \ee
in agreement with our consistent results in the previous section,
\eqref{central charges1}.

\section{Relation to 3D gravity}

In this section we would like to compare our 2D solutions with
those in 3D  Einstein-Chern-Simons gravity which have recently
been studied in \cite{strominger2}. To do so, we note that the two
dimensional $AdS_2$ solutions \eqref{solutions} may be uplifted to
three dimensions.

In general if we start from a 2D solution
 \be
 ds_2^2=g_{\mu\nu}dx^\mu dx^\nu,\;\;\;\;\;e^{\phi},\;\;\;\;\;A_\mu,
 \ee
which we assume to be symmetric under an isometry group ${\cal
G}$, we can find a pure geometric 3D gravity solution
 \be
 ds^2_3=e^{2\phi}\bigg{[}ds_2^2+\bigg{(}dy+l A_\mu
 dx^\mu\bigg{)}^2\bigg{]}
 \ee
with isometry ${\cal G}\times U(1)$. Here $y$ is a coordinate that
parameterizes an $S^1$ with period $2\pi l$.

In particular consider the case where the two dimensional solution
is $AdS_2$. The isometry group of the solution is
$SL(2,R).$\footnote{The solution may have extra symmetries. For
example for the solutions \eqref{solutions} we have $U(1)$ gauge
symmetry as well.} Being symmetric under $SL(2,R)$ group the
solution has constant dilaton and $F_{tr}$. By uplifting the
solution to three dimensions we find a pure geometric solution whose
isometry is $SL(2,R)\times U(1)$; the obtained solution will be
$S^1$ fibered over $AdS_2$. In other words, in light of the recent
terminology, the solution may be thought of as {\it warped} $AdS_3$
\cite{strominger2}. For particular values of the radius of the
$AdS_2$ space and field strength, the resultant three dimensional
solution describes a locally $AdS_3$ solution. However globally it
is $AdS_3$ with an identification. The effect of this identification
is that the isometry group of $AdS_3$, $SL(2,R)\times SL(2,R)$,
breaks to $SL(2,R)\times U(1)$, as mentioned above.

Applying the above procedure to the solutions \eqref{solutions} we get
\bea\label{3dsolutions}
1: && ds^2=\frac{l^2}{4}\left(-r^2dt^2+\frac{dr^2}{r^2}+(dz-rdt)^2\right) ,
\;\;\;\;\;\;\;\;\;\;\;\;\;\;\;\;\;\;\;\;\;\;\;\;\;\;\;\;\;q<0,\cr &&\cr
2: && ds^2=\frac{l^2}{4}\left(-r^2dt^2+\frac{dr^2}{r^2}+(dz+rdt)^2\right) ,
\;\;\;\;\;\;\;\;\;\;\;\;\;\;\;\;\;\;\;\;\;\;\;\;\;\;\;\;\;q>0,\cr &&\cr
3: &&ds^2=\frac{9l^2}{\mu^2
l^2+27}\left(-r^2dt^2+\frac{dr^2}{r^2}+\frac{4\mu^2 l^2}{\mu^2l^2+27}(dz+rdt)^2\right),\;\;\;\;q>0,
\eea
where $z=l\theta/|e|$ with the identification $z\sim z+ 2\pi l n \frac{l}{|e|}$. Here $n$ is an integer.

We note, however, that the above description must be considered with special care. It is known that
the asymptotic symmetry of the $AdS_2$ is is a copy of the Virasoro algebra whose global part is an $SL(2,R)$ \cite{Strominger:1998yg}. This is, indeed, the generalization of $AdS_3$ where the asymptotic symmetry
is two copies of
the Virasoro algebra with $SL(2,R)_L\times SL(2,R)_R$ global part \cite{BH}.
It is crucial to note that, in general, the global part of the Virasoro algebra of $AdS_2$ geometry is not
necessarily the $SL(2,R)$ symmetry which only leaves the metric invariant.
Indeed as we have seen in the previous section
the asymptotic symmetry of $AdS_2$ solutions of \eqref{solutions} is given by the twisted energy momentum
tensor. Now uplifting the solutions to three dimensions the resultant $SL(2,R)$ must be read from
the twisted energy momentum tensor. In other words, if we denote the left/right handed energy momentum tensor
of the three dimensional theory by $T^{(3)}_{\pm\pm}$, one should identify
$T^{(3)}_{\pm\pm}={\tilde T}_{\pm\pm}$ \cite{Strominger:1998yg}.

Since in two dimensions the
theory is chiral, upon uplifting the theory to three dimensions we only get non-zero excitations
for one hand. In other
words, depending on whether the two dimensional
solution is left/right handed we will have left/right handed three dimensional energy momentum tensor.
Actually from 3D point of view, as we have already mentioned, due to the identification the excitation
states live purely in  $SL(2,R)_L$or $SL(2,R)_R$ factor of the isometry group.

On the other hand as we have seen in the previous section the 2D
twisted energy momentum tensor has non-zero central charge given
by \eqref{central charges1}. Therefore the corresponding central
charge of the dual CFT of the three dimensional solutions is given
by \be\label{c1} 1:\;\;c_R=\frac{3l}{2G_3}(1+\frac{1}{\mu
l}),\;\;\;\;\;\; 2:\;\;c_L=\frac{3l}{2G_3}(1-\frac{1}{\mu
l}),\;\;\;\;\;\; 3:\;\;c_L=\frac{12\mu l^2}{G_3(\mu^2 l^2+27)},
\ee where $G_3$ is 3D Newton constant. Of course, although the
theory we get has excitations of only one hand, the other sector
exists but has zero excitations. Thus the 2D CFT dual to the above
3D solutions has both $c_L$ and $c_R$. Using the diffeomorphism
anomaly by which we have $c_L-c_R=-3/\mu G_3$ \cite{KL}, one finds
\be\label{c2} 1:\;\;c_L=\frac{3l}{2G_3}(1-\frac{1}{\mu
l}),\;\;\;\;\;\; 2:\;\;c_R=\frac{3l}{2G_3}(1+\frac{1}{\mu
l}),\;\;\;\;\;\; 3:\;\;c_R=\frac{15\mu^2 l^2+81}{\mu G_3(\mu^2
l^2+27)}, \ee in agreement with \cite{sen2} and
\cite{strominger2}.

As we have already mentioned the $z$ coordinate in solutions \eqref{3dsolutions} is periodic. Therefore
one may interpret the solutions as 3D extremal black holes. Due to the identification
the $SL(2,R)_L/SL(2,R)_R$-invariant $AdS_3$ vacuum should give a thermal state for the left/right movers
of the boundary CFT with zero right/left temperature and non-zero left/right temperature. On the other hand
the $t$ direction
can be treated as the null coordinate of the boundary, while the $z$ should be considered as
a Rindler coordinate. Therefore the left/right temperature of the dual CFT is
proportional to the magnitude of the shift in $z$ direction. More precisely, one gets\footnote{Note that
such a treatment for warped $AdS_3$ is tricky due to its boundary. Nevertheless in writing the expression for
this case we are encouraged by the fact that in this case the picture fits nicely as well.}
 \be
1:\;T_R=\frac{2l}{\pi} \sqrt{\frac{-qG_3}{l(1+\frac{1}{\mu l})}},\;\;
2:\;T_L=\frac{2l}{\pi} \sqrt{\frac{qG_3}{l(1-\frac{1}{\mu l})}},\;\;
3:\;T_L=\frac{2l}{\pi}\sqrt{\frac{qG_3(\mu^2 l^2+27)}{8\mu l^2}}.
\ee
The corresponding entropy using the Cardy formula $S=\frac{\pi^2}{3}c_{L/R} T_{L/R}$ reads

\bea
1:&&S=2\pi\sqrt{\frac{-ql^2}{6}\frac{3l}{2G}(1+\frac{1}{\mu l})},\cr
2:&&S=2\pi\sqrt{\frac{ql^2}{6}\frac{3l}{2G}(1-\frac{1}{\mu l})},\cr
3:&&S=2\pi \sqrt{\frac{ql^2}{6}\frac{12\mu l^2}{G(\mu^2l^2+27)}},
\eea
which are compatible with those we have found in section two from 2D point of view.

\section{Conclusions}

In this paper we have studied 2D Maxwell-dilaton gravity on $AdS_2$
geometry in the presence of higher order correction given by
Chern-Simons term. The model admits three distinctive $AdS_2$ vacuum
solutions characterized by the sign of the electric field. Using the
entropy function formalism we have evaluated the entropy of the
solutions. Note that in the leading order when the action is given
by the Einstein-Hilbert action the model has only one solution.
Adding the Chern-Simons term the solution gets corrections which
depend on the coefficient of the Chern-Simons term as well as the
sign of the electric charge leading to three different solutions.
The sign dependent nature of the corrections may be associated with
the fact that the dual CFT is believed to be chiral half of a 2D
CFT.

When the coefficient of the Chern-Simons term is set to zero the solution (1) degenerates with
solution (2) while the third one disappears. In other words, the solutions (1) and (2) are
Einstein solutions while the last one is not. Of course for particular values of $\mu l$
the third one degenerates with the solution (2) as well.

Following \cite{HS} we have studied the action of the conformal
group in the model where we have seen that in order to have
consistent boundary conditions we will have to work with a twisted
energy momentum tensor. The twisted energy momentum tensor has
non-zero central charge which should be associated with the central
charge of the dual CFT. Requiring to have a consistent picture we
have been able to read the corresponding central charge as well as
the level of $U(1)$ current.

We have  compared our solutions with those in 3D gravity by uplifting the solutions to three
dimensions. The solutions (1) and (2) have been uplifted to a solution which is locally
$AdS_3$ though globally it is $AdS_3$ with an identification. The third one has been uplifted
to a solution which is known as warped $AdS_3$ \cite{strominger2} with an identification.
Due to the identification, the resultant 3D solutions may be thought of as 3D extremal black holes.

We have also determined the entropy of the extremal black holes which are given by the Cardy
formula using the obtained central charges. The consistency of the results points toward the
conjecture made in \cite{strominger2} where the authors proposed that the 3D gravity on the
warped $AdS_3$  geometry is dual to a 2D CFT with
left and right hand central charges given by the third central charges in equations
\eqref{c1} and \eqref{c2}.

The $AdS_3$ and warped $AdS_3$ solutions in 3D gravity are believed
to be dual to 2D CFTs with $c_L$ and $c_R$ given by equations
\eqref{c1} and \eqref{c2}. Therefore in general one might expect
that from two dimensional point of view we should have got four
solutions corresponding to four different sectors which are obtained
from 3D $AdS_3$ and warped $AdS_3$. But as we have seen in two
dimensions only three of them can be realized. The missing one is
the right handed sector of the warped $AdS_3$ solution with central
charge given by the third one of \eqref{c2}. This means that if we
consider an extremal black hole in warped $AdS_3$ there is only one
possibility in which the left movers will survive. This is unlike an
extremal black hole in $AdS_3$ where it could be either left or
right handed with non-zero excitations of left or right mover
states, respectively.

It is worth mentioning that, as it was observed by the authors of
\cite{compere}, when we study asymptotic symmetry of the warped
$AdS_3$ solution one gets a copy of Virasoro algebra with central
charge given by the third one of \eqref{c2}. This is exactly the one
which cannot be realized from 2D point of view. It would be
interesting to illustrate the physics behind this special behavior
of the warped $AdS_3$.

It was shown in \cite{LSS} that the TMG quantum mechanically makes
sense only at $\mu l=1$ where we get chiral gravity. From 2D point
of view although we have observed that $\mu l=1$ is a special point,
it is not a priori clear why we should set $\mu l=1$ from 2D point
of view. Indeed, there are several examples in string theory where
we have extremal black holes which upon reduction to two dimensions
we get an action very similar to that in \eqref{actions}. In these
cases the coefficient of the Chern-Simons term usually is fixed by a
topological number and the charges of black holes. As far as the
black holes are concerned there are no conditions on the coefficient
of the Chern-Simons term. It would be interesting to understand this
point better

\vspace*{1cm}

{\bf Acknowledgments}

We would like to thank Farhad Ardalan for useful and illustrative discussions on $AdS_2/CFT_1$ correspondence.
This work is supported in part by Iranian TWAS chapter at ISMO.


\begin{thebibliography}{99}


\bibitem{witten}
  E.~Witten,
  ``Three-Dimensional Gravity Revisited,''
  arXiv:0706.3359 [hep-th].



\bibitem{Banados:1992wn}
  M.~Banados, C.~Teitelboim and J.~Zanelli,
  ``The Black hole in three-dimensional space-time,''
  Phys.\ Rev.\ Lett.\  {\bf 69}, 1849 (1992)
  [arXiv:hep-th/9204099].


\bibitem{Maldacena:1997re}
  J.~M.~Maldacena,
  ``The large N limit of superconformal field theories and supergravity,''
  Adv.\ Theor.\ Math.\ Phys.\  {\bf 2}, 231 (1998)
  [Int.\ J.\ Theor.\ Phys.\  {\bf 38}, 1113 (1999)]
  [arXiv:hep-th/9711200].



\bibitem{Strominger:1998yg}
  A.~Strominger,
  ``AdS(2) quantum gravity and string theory,''
  JHEP {\bf 9901}, 007 (1999)
  [arXiv:hep-th/9809027].

\bibitem{Cadoni:1999ja}
  M.~Cadoni and S.~Mignemi,
  ``Asymptotic symmetries of AdS(2) and conformal group in d = 1,''
  Nucl.\ Phys.\  B {\bf 557}, 165 (1999)
  [arXiv:hep-th/9902040].

\bibitem{NavarroSalas:1999up}
  J.~Navarro-Salas and P.~Navarro,
  ``AdS(2)/CFT(1) correspondence and near-extremal black hole entropy,''
  Nucl.\ Phys.\  B {\bf 579}, 250 (2000)
  [arXiv:hep-th/9910076].

\bibitem{Park:1999hs}
  M.~I.~Park and J.~H.~Yee,
  ``Comments on 'Entropy of 2D black holes from counting microstates',''
  Phys.\ Rev.\  D {\bf 61}, 088501 (2000)
  [arXiv:hep-th/9910213].



\bibitem{Cadoni:2000ah}
  M.~Cadoni and S.~Mignemi,
  ``Symmetry breaking, central charges and the AdS(2)/CFT(1)  correspondence,''
  Phys.\ Lett.\  B {\bf 490}, 131 (2000)
  [arXiv:hep-th/0002256].


\bibitem{Cadoni:2000gm}
  M.~Cadoni, P.~Carta, D.~Klemm and S.~Mignemi,
  ``AdS(2) gravity as conformally invariant mechanical system,''
  Phys.\ Rev.\  D {\bf 63}, 125021 (2001)
  [arXiv:hep-th/0009185].


\bibitem{Astorino:2002bj}
  M.~Astorino, S.~Cacciatori, D.~Klemm and D.~Zanon,
  ``AdS(2) supergravity and superconformal quantum mechanics,''
  Annals Phys.\  {\bf 304}, 128 (2003)
  [arXiv:hep-th/0212096].

\bibitem{Kim:1998wy}
  W.~T.~Kim,
  ``AdS(2) and quantum stability in the CGHS model,''
  Phys.\ Rev.\  D {\bf 60}, 024011 (1999)
  [arXiv:hep-th/9810055].

\bibitem{Leiva:2003kd}
  C.~Leiva and M.~S.~Plyushchay,
  ``Conformal symmetry of relativistic and nonrelativistic systems and  AdS/CFT
  correspondence,''
  Annals Phys.\  {\bf 307}, 372 (2003)
  [arXiv:hep-th/0301244].
  
\bibitem{Hyun:2007ii}
  S.~Hyun, W.~Kim, J.~J.~Oh and E.~J.~Son,
  ``Entropy Function and Universal Entropy of Two-Dimensional Extremal Black
  Holes,''
  JHEP {\bf 0704}, 057 (2007)
  [arXiv:hep-th/0702170].
  
\bibitem{Correa:2008bi}
  F.~Correa, V.~Jakubsky and M.~S.~Plyushchay,
  ``Aharonov-Bohm effect on $AdS_2$ and nonlinear supersymmetry of reflectionless
  Poschl-Teller system,''
  arXiv:0809.2854 [hep-th].






\bibitem{Kang:2004js}
  G.~Kang, J.~i.~Koga and M.~I.~Park,
  ``Near-horizon conformal symmetry and black hole entropy in any  dimension,''
  Phys.\ Rev.\  D {\bf 70}, 024005 (2004)
  [arXiv:hep-th/0402113].




\bibitem{HS}
  T.~Hartman and A.~Strominger,
  ``Central Charge for $AdS_2$ Quantum Gravity,''
  arXiv:0803.3621 [hep-th].


\bibitem{Sen:2008yk}
  A.~Sen,
  ``Entropy Function and $AdS_2/CFT_1$ Correspondence,''
  arXiv:0805.0095 [hep-th].


\bibitem{Alishahiha:2008tv}
  M.~Alishahiha and F.~Ardalan,
  ``Central Charge for 2D Gravity on AdS(2) and AdS(2)/CFT(1) Correspondence,''
  JHEP {\bf 0808}, 079 (2008)
  [arXiv:0805.1861 [hep-th]].

\bibitem{Gupta:2008ki}
  R.~K.~Gupta and A.~Sen,
  ``Ads(3)/CFT(2) to Ads(2)/CFT(1),''
  arXiv:0806.0053 [hep-th].

\bibitem{Cadoni:2008mw}
  M.~Cadoni and M.~R.~Setare,
  ``Near-horizon limit of the charged BTZ black hole and $AdS_2$ quantum
  gravity,''
  arXiv:0806.2754 [hep-th].



\bibitem{Castro:2008ms}
  A.~Castro, D.~Grumiller, F.~Larsen and R.~McNees,
  ``Holographic Description of AdS$_2$ Black Holes,''
  arXiv:0809.4264 [hep-th].



\bibitem{Sen:2008vm}
  A.~Sen,
  ``Quantum Entropy Function from AdS(2)/CFT(1) Correspondence,''
  arXiv:0809.3304 [hep-th].



\bibitem{Morita:2008qn}
  T.~Morita,
  ``Hawking Radiation and Quantum Anomaly in AdS2/CFT1 Correspondence,''
  arXiv:0811.1741 [hep-th].


\bibitem{SEN}
  A.~Sen,
  ``Black hole entropy function and the attractor mechanism in higher
  derivative gravity,''
  JHEP {\bf 0509}, 038 (2005)
  [arXiv:hep-th/0506177].



\bibitem{strominger2}
  D.~Anninos, W.~Li, M.~Padi, W.~Song and A.~Strominger,
  ``Warped $AdS_3$ Black Holes,''
  arXiv:0807.3040 [hep-th].


\bibitem{compere}
  G.~Compere and S.~Detournay,
  ``Semi-classical central charge in topologically massive gravity,''
  arXiv:0808.1911 [hep-th].



\bibitem{Anninos:2008qb}
  D.~Anninos,
  ``Hopfing and Puffing Warped Anti-de Sitter Space,''
  arXiv:0809.2433 [hep-th].

\bibitem{Carlip:2008eq}
  S.~Carlip, S.~Deser, A.~Waldron and D.~K.~Wise,
  ``Topologically Massive AdS Gravity,''
  Phys.\ Lett.\  B {\bf 666}, 272 (2008)
  [arXiv:0807.0486 [hep-th]].

\bibitem{Gibbons:2008vi}
  G.~W.~Gibbons, C.~N.~Pope and E.~Sezgin,
   ``The General Supersymmetric Solution of Topologically Massive
  Supergravity,''
  Class.\ Quant.\ Grav.\  {\bf 25}, 205005 (2008)
  [arXiv:0807.2613 [hep-th]].





\bibitem{Guralnik:2003we}
  G.~Guralnik, A.~Iorio, R.~Jackiw and S.~Y.~Pi,
  ``Dimensionally reduced gravitational Chern-Simons term and its kink,''
  Annals Phys.\  {\bf 308}, 222 (2003)
  [arXiv:hep-th/0305117].


\bibitem{Grumiller:2003ad}
  D.~Grumiller and W.~Kummer,
  ``The classical solutions of the dimensionally reduced gravitational
  Chern-Simons theory,''
  Annals Phys.\  {\bf 308}, 211 (2003)
  [arXiv:hep-th/0306036].


\bibitem{sen2}
  B.~Sahoo and A.~Sen,
  ``BTZ black hole with Chern-Simons and higher derivative terms,''
  JHEP {\bf 0607}, 008 (2006)
  [arXiv:hep-th/0601228].



\bibitem{Manton:1985jm}
  N.~S.~Manton,
  ``The Schwinger Model And Its Axial Anomaly,''
  Annals Phys.\  {\bf 159} (1985) 220.

\bibitem{Heinzl:1991vd}
  T.~Heinzl, S.~Krusche and E.~Werner,
  ``Nontrivial vacuum structure in light cone quantum field theory,''
  Phys.\ Lett.\  B {\bf 256} (1991) 55
  [Nucl.\ Phys.\  A {\bf 532} (1991) 429C].



\bibitem{BH}
  J.~D.~Brown and M.~Henneaux,
   ``Central Charges in the Canonical Realization of Asymptotic Symmetries: An
  Example from Three-Dimensional Gravity,''
  Commun.\ Math.\ Phys.\  {\bf 104}, 207 (1986).



\bibitem{KL}
P. Kraus and F. Larsen, ``Holographic gravitational anomalies,''
JHEP  {\bf 0601}, 022 (2006) [arXiv:hep-th/0508218].



\bibitem{LSS}
  W.~Li, W.~Song and A.~Strominger,
  ``Chiral Gravity in Three Dimensions,''
  JHEP {\bf 0804}, 082 (2008)
  [arXiv:0801.4566 [hep-th]].





\end{thebibliography}
\end{document}